# Exploring Optical and Electrical Gas Detection Based on Zinc-Tetraphenylporphyrin Sensitizer

Gulimire Tuerdi*, Abliz Yimit**, Xiaoyan Zhang*†

*School of Materials Science and Engineering, Tsinghua University, Beijing 100084, China.
**College of Chemistry and Chemical Engineering, Xinjiang University, Urumqi 830046, China.
†To whom correspondence should be addressed
E-mail: xyzhang@iccas.ac.cn

We developed optical waveguide (OWG), ultraviolet-visible spectrophotometry (UV-vis), and electrically operated gas sensors by utilizing zinc-tetraphenylporphyrin (ZnTPP) as sensitizer. Strikingly, ZnTPP thin-film/$K^+$-exchanged glass optical waveguide (OWG) sensing element exhibits a superior signal-to-noise ratio (SNR) of 109.6 upon 1 ppm $NO_2$ gas injection, which is 29.5 and 3.8 times larger than that of UV-vis (absorbance at wavelength of 438 nm) and ZnTPP electrical sensing elements prepared on an alumina ceramic tube, respectively. Further results on Fourier Infrared spectra (FT-IR) and UV-vis spectra, confirm a strong chemical adsorption of $NO_2$ gas on ZnTPP. Therefore, our studies highlight the selection of suitable detection technique for analyte sensing with ZnTPP.

**Introduction**

Harmful and toxic gases have serious effects on human health, air quality and public safety. Recently increased serious environmental problems, such as acid rain and photo chemical smog, were mainly caused by increased $NO_2$ gaseous anthropogenic emissions from different industrial process (i.e. heating, power generation and engines, etc.) as well as automobile exhaust gaseous.[1] It is worth noting that even parts per million (ppm) level of $NO_2$, as an air pollutant, can cause symptoms of reduced lung function growth and bronchitis in asthmatic children increase in association with long-term exposure.[2] In this respect, it is an urgent need for designing a room temperature $NO_2$ gas senor with high selectivity as well as high sensitivity. In recent years, semiconducting metal oxides based gas sensors (for example ZnO, $SnO_2$, CuO, $In_2O_3$, NiO, and $WO_3$) have attracted wide attention due to their excellent performance, such as high sensitivity and repeatability.[3-5] A $NO_2$ gas sensor based on 3D hierarchical monoclinic-type structural Sb-doped $WO_3$ is reported exhibits a detection limit of 1 ppm at 30°C and 0.5 ppm at 70°C, respectively.[6] However, such semiconducting metal oxide based sensors generally have critical disadvantages, such as require high operating temperatures, relatively high cost, and complexed preparation process, which limit their wide applications including on-board gas sensors (particularly in the large motor-vehicle market environment). Therefore, it is highly demanded for sensitive and low-cost sensing materials that can specifically bind to target analysts.

Porphyrins as natural pigments, are ubiquitous in nature and perform essential functions of life.[7] Generally, porphyrins consist of four pyrrole rings connected by methines. Open structure of porphyrins allows easy access for gas molecules, which can interact with π-electrons in porphyrins.[8] When porphyrins coordinated with metal complexes (such as zinc, copper, nickel, cobalt or iron) and parts of many proteins with various functions, the large π-aromatic system on porphyrins can lead to interesting optical properties as well as high optical absorption coefficient.[9, 10] Furthermore, porphyrins and related metallo-porphyrins exposed to analyte gases are able to occur prominent opto-electrical changes.[11] Therefore, porphyrins and its metallo-porphyrins have been considered as prospective gas sensing materials for optical gas sensors.[12]

Detection of gas molecules by optical sensing techniques is a hot topic of research interest. Thin-film based glass OWG gas transducer consists of two key principles: the absorbance intensity of the thin-film is directly affected by its interaction with the analyte gases, and the intensity of output light from the glass OWG thin-film is also related to the absorbance of the transducer due to the interaction with analyte gases. Since 1990s, scientists have paid great attention to researches and applications of OWG transducers in the field of optical communication.[13, 14] Thin-film glass OWG transducers have been utilized for fluorescence and/or electronic absorbance detection of toxic agents as well as clinical analytes.[15] The OWG analyte detection method has many merits over the other types of sensors, such as fast response/recovery times, remote controllability, potentially high sensitivity, intrinsically safe detection, and room temperature operation.[16-18]

Selectivity and sensitivity are two of the most important factors to evaluate the performance of a gas sensor. Although researchers have developed several kinds of ultra-sensitive $NO_2$ gas sensors based on various materials, analysis and comparation on a certain gas sensitive material with different detection methods to evaluate the best is rarely so far. Herein, we studied a highly sensitive ZnTPP film for detection of low concentration of $NO_2$ gas at room temperature by utilizing OWG technique, UV-vis spectrum and electrical resistive measurement, respectively. We also discussed the mechanism for highly sensitive $NO_2$ gas sensing mechanism of the sensor based on ZnTPP film.

**Experimental**

**Preparation of testing gases**

Standard $NO_2$ gas was obtained by reacting a given amount of Cu with $HNO_3$ at room temperature and normal pressure in a 600 mL standard vessel. The concentration of as prepared gas was confirmed using commercial gas detection tubes (working range of 2-200 ppm, Gastec, Beijing Municipal Institute). Different amounts of standard $NO_2$ gas were diluted with dry air in a 600 mL second standard vessel to obtain the desired concentrations. Utilizing this standard vessel-dilution method, a very low concentration of $NO_2$ (10 ppb) could be obtained.

**Gas sensing elements**

**Planar OWG sensing device**

The planar OWG sensing element consists of a glass substrate, $K^+$ ion exchanged layer, a sensing layer and covering material (usually air). The 0.05% ZnTPP solution was spin-coated on glass substrates (76 mm×26 mm×1 mm, microscopy slide, Citotest Labware Manufacturing Co., Ltd, China) at various rotation speed from 800 to 1050 rpm for 30 min. Before coating, the glass substrates (composed of $SiO_2$, $Na_2O$, and CaO) was subjected to a K$^+$-exchanged process by utilizing thermal ion-exchange method.[19] Glass slide was immersed into the molten KNO$_3$ at 400 °C for 40 min. Ionic polarizability, molar volume and stress state created by the substitution decide the net-index of the system. In the K$^+$-Na$^+$ ion-exchange method, the polarizability of K$^+$ ion is slightly higher than that of Na$^+$ ion on glass. Therefore, K$^+$ ion can be incorporated into the glass (~1-2 μm deep), and substantially result in index change lower (n≤0.01) along with a smaller diffusion rate. Afterwards, the K$^+$-ion exchanged samples were washed with deionized water, ethanol and acetone, respectively. Finally, the coated elements were heated at 60 °C for 10 min to induce desorption of any solvent in the film before placed in a vacuum drying oven for overnight at room temperature. The schematic illustration of the ZnTPP film-thin/K$^+$-exchanged glass OWG device is shown in Fig. 1a.

After the drying process, a dielectric laser (λ=650 nm) beam was introduced to the OWG utilizing a prism (refractive index, *n*, 1.78) coupler, which was emerged from another prism coupler. Then, diiodomethane is used as a matching liquid (*n*=1.74) between glass slide and prism. Two prisms are symmetrically placed on OWG gas sensing element to have the sensitive layer in between, as shown in Fig. 1a. The width of the sensitive layer was 3 mm, and the distance between sensitive layer and prisms was 5 mm.

In the TE$_0$ propagation mode, the ZnTPP film-thin/K$^+$-exchanged waveguide on a glass-substrate was described with Rung-Kutta method expressed by the following equation:[20]

$$S_{OWG} = \left(\frac{n_{surf}^2}{2N_{eff}}\right) \frac{E_y(0)^2}{\int_{-\infty}^{+\infty} E_y(x)^2 dx} \quad (1)$$

where, $S_{OWG}$ donates the sensitivity of the ZnTPP film glass OWG sensing element, and $N_{eff}$ and $n_{surf}$ are the effective index and surface refractive index of the waveguide, respectively. Wherein, $n_{surf}$ can be expressed by $(n_f^2+n_c^2)^{1/2}$, where $n_c$ and $n_f$ donate the refractive index of the cladding layer and the K$^+$-exchanged glass ZnTPP film, respectively. Ey(0) and Ey(x) donate the electric field intensity on the OWG surface and electric filed distribution of the guided light.

**OWG gas sensing measurement:** gas sensing measurements of ZnTPP film-thin/K$^+$-exchanged glass OWG sensing element were performed utilizing a homemade OWG detection system. The gas sensing apparatus is similar to that described in previous report,[17] which is consisted of ZnTPP film OWG sensor, a dielectric laser beam (650 nm), a compressed air source, a flow meter, a diffusion tube, a reflector, a light detector, and a computer as a recorder. The K$^+$-exchanged ZnTPP film OWG sensing element was attached to the prism in an index matching liquid, and fixed to the gas measurement cell. The intensity of the output light was detected by light detector and output signal was recorded by a recorder, respectively. For each gas detection measurement, a syringe of 20 cm$^3$ fresh gas sample was injected into the flow chamber, which was then vented out. Dry air was directed through the chamber at a constant rate of 50 cm$^3$ min$^{-1}$, in order to carry each sample gas to the sensitive layer.

**UV-vis gas sensing measurement**: gas sensing measurement of the ZnTPP film sensor was performed using optical absorption spectroscopy (UV-1780, Shimadzu Technology Co. Ltd, China). For each measurement, a new syringe and cuvette were used for detection of different concentration of gases. Numbers of films, prepared under optimum condition, were cut into width of 3 cm and placed in a cuvette to be detected. The sensitivity of the ZnTPP UV-vis film sensing element was expressed by $\Delta Abs$ ($Abs_{air}$-|$Abs_{gas}$|), where $Abs_{gas}$ and $Abs_{air}$ were absorbance of the sensor in air and mixture of testing gas in air, respectively.

**Electrical resistive gas sensing measurement:** the resistive gas sensing property of the ZnTPP gas sensing element, a ring-shape four-electrode conductance sensor, was examined in a cubical chamber with the volume of $20 \times 25 \times 40$ cm$^3$. The resistance of the sensor was examined by a conventional electrical circuit, in which the sensor was linked with an external resistor in series at a circuit voltage of 5 V. The response of the sensitive element was determined to be $R_{air}/R_{gas}$, where $R_{air}$ donates the resistance of the sensitive element in air and $R_{gas}$ donates in the mixture of the analyte gas in air. The whole operation process was carried out in the clean room. Pressure, temperature, and humidity of the clean room were one atmospheric pressure, 25°C, and 25% RH, respectively.

**Instruments**

Surface morphology and film thickness of the scanning electron microscopy (SEM, XL-30E, Philips, Germany). Optical absorption behavior of ZnTPP solution and thin-film was characterized by electronic absorption spectra (UV-1780, Shimadzu Technology Co. Ltd, China). Gas sensing measurements were operated by home-made OWG gas sensing apparatus, UV-vis absorption spectra, and four-electrode electrical conductance sensing measurement apparatus, respectively. Gas sensing mechanism of the sensing device before and after exposure to analyte gas was tested by utilizing emission (FTIR-650, Beijing, China) and UV-vis spectra.

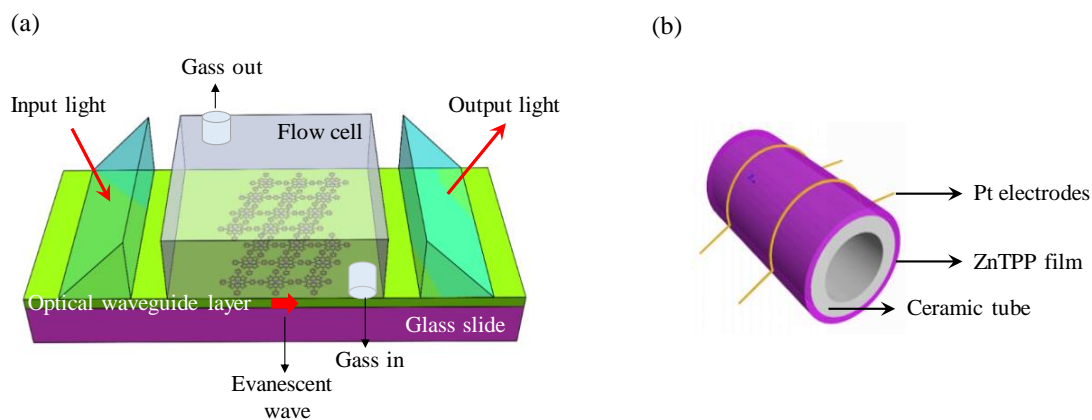

Fig. 1 Schematic illustration of two sets up for gas sensors. (a) OWG film gas sensing element, and (b) electrical resistive gas sensing element.

## Results and Discussion

### Morphology and sensing characteristics

The UV-vis spectra of 4.75 uM ZnTPP solution in chloroform and solid ZnTPP thin-film (2.56 mM, spinning coating rate of 1050 rpm for 30 s) are given in Fig. 2. The spin-coated ZnTPP thin-film appeared broaden UV-vis characteristic peak (i.e. full width at half-maximum of 25 nm for Soret-band and 10 nm for Q-band) and red-shifted compared to the ZnTPP solution. The observed broadening and red-shifting of the Soret-band in the ZnTPP thin-film indicate that a certain number of ZnTPP molecules formed J-aggregates (i.e. head-to-tail arrangement).[21] The ZnTPP aggregates can be attributed to the occurrence of an incoherent interaction in the surroundings of the dense film.[21, 22]

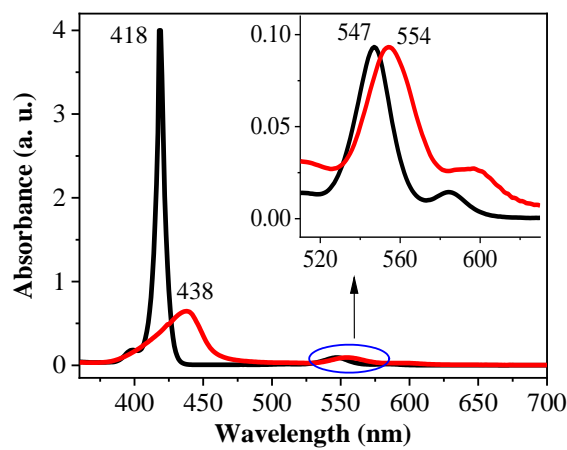

Fig. 2 UV-vis spectra of ZnTPP solution in chloroform (black curve) and solid thin-film (red curve).

Intermolecular π-π interaction between porphyrin rings can cause head-to-tail aggregation effect, which is detrimental to sensitivity of the sensing device by preventing the porphyrin-analyte interaction.[23] Therefore, aggregation degree of ZnTPP molecules on the glass substrate is one of the major factor for determination of the sensing capability of the device. In order to find out the optimal fabrication condition for ZnTPP thin-film with minimal aggregates, we spin-coated porphyrin solutions in chloroform with various mass concentrations (0.05%, 0.07%, 0.09%, 0.12%, and 0.15%) on glass slides under different rotating speeds. SEM images show under the same ZnTPP concentration, higher rotating speed resulting less aggregations (Fig. 3abc), while under the same rotating speed both low concentration of 0.09% (Fig. 3d) and high concentration of 0.15 (Fig. 3f) show high degree of aggregations. Therefore, 0.12% and 1050 rpm at 30 s were the optimum condition for fabricating ZnTPP thin-film $K^+$-exchanged glass OWG for sensing application. Under this condition, porphyrin film depicts nanorod aggregates with average length of 100-130 nm and width of ~20 nm for single nanorod, and distance of ≥100 nm between ZnTPP nanorods, respectively (Fig. 3e). According to the previous studies, nanorod aggregation distance in range of nanometer to micrometer shows optimal sensitivity.[24] And in a distance of ≥100 nm, at least ≥200 $NO_2$ gas molecules can entrance into the ZnTPP nanorod gaps and react with porphyrin molecules. In the following studies, we mainly used the ZnTPP thin-film that fabricated under concentration of 0.12% and rotation speed of 1050 rpm at 30 s, otherwise we stated. Such obtained solid ZnTPP thin-film shows a thickness about 51 nm measured by SEM (Fig. S1), which is in good agreement with theoretical calculation[25] and previous studies on optimal sensing film thickness. It was reported that the sensitivity of the ZnTPP film OWG reached its top value with a mean thickness in the range of 50-55 nm for dielectric laser beam (at wavelength of 650 nm), which could support the guiding mode of the transverse electric (TE$_0$) mode.[26]

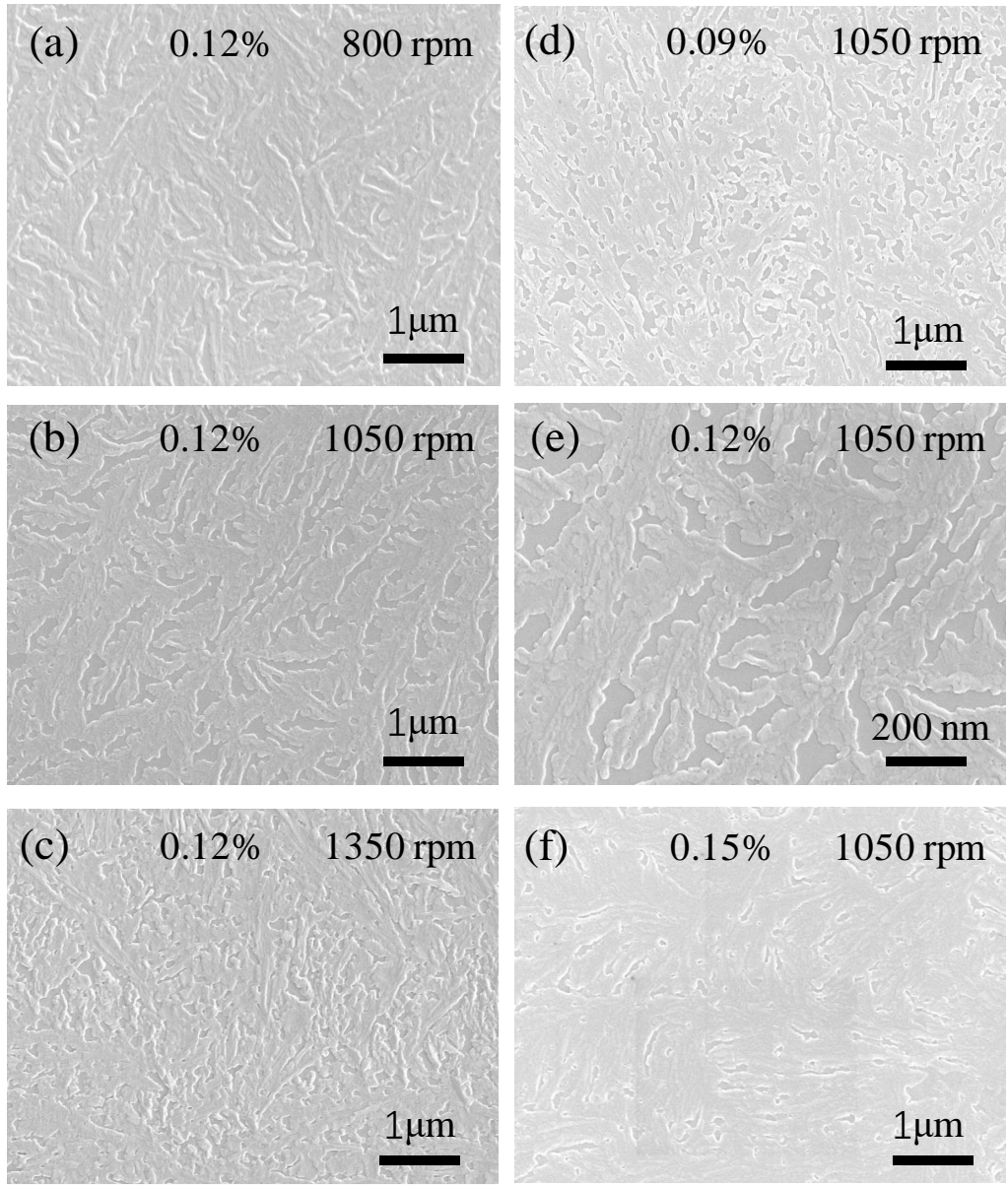

Fig. 3 SEM images of the ZnTPP thin-films under different mass percentage concentrations and rotating speeds.

ZnTPP thin-film coated on a glass slide was exposed to 100 ppm of $NO_2$ gas, and UV-vis absorbance at wavelength of 300-750 nm was monitored. The typical absorption spectrum of ZnTPP film contains Soret band at 437 nm and Q band at 555 nm before exposure to $NO_2$ gas (Fig. 4 black curve), which show a significant change after exposure to $NO_2$ gas. Specifically, there is obvious decrease in intensity of the absorption peaks at both Soret and Q bands of ZnTPP film (Fig. 4 red curve). In addition, the color of the ZnTPP thin-film visibly changed from light yellow to yellow green (show as insert in Fig. 4), resulting in the intensity of output light reduce.

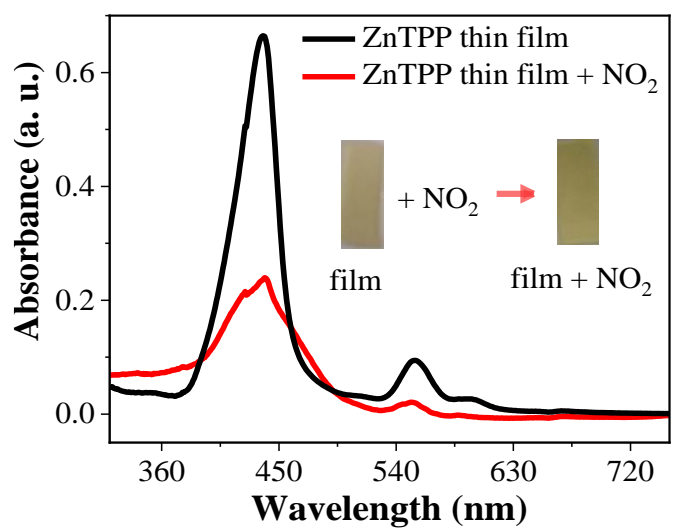

Fig. 4 UV-vis absorbance spectra and color change of ZnTPP thin-film before and after exposure to 100 ppm $NO_2$ gas.

### Gas sensing performance
*OWG sensing*

For ZnTPP thin-film/$K^+$-exchanged glass OWG gas sensing, the device was placed in OWG gas measurement system for $NO_2$ gases test. During the whole process, dry air was flowed throughout and functioned as dilute and carrier gas. As the ZnTPP thin-film exposed to $NO_2$ gas, the output light intensity decreased steadily. Signal intensity change ($\Delta I$) was calculated by $\Delta I = I_{gas} - I_{air}$, wherein, $I_{air}$ and $I_{gas}$ are intensities of output light of ZnTPP thin-film/$K^+$-exchanged glass OWG sensing element before and after exposure to analyte gases, respectively. The intensity of output light was decreased in the presence of $NO_2$ gas in the 10 ppb -100 ppm range (Fig. 5a). Particularly, OWG sensor reveals good response ($\Delta I = 61$) to $NO_2$ gas concentration as low as 10 ppb, with a signal-to-noise ratio (SNR) of 3.4 and short response time of <1 s. A calibration curve of the sensing signal with the NO$_2$ gas concentration was obtained by plotted $\Delta I$ of the sensing element against the concentration of NO$_2$ gas. The response of the ZnTPP thin-film to NO$_2$ gas was strongly dependent on NO$_2$ gas concentration, and demonstrated linear relationship in the NO$_2$ concentration range of 10 ppb - 100 ppm ($\Delta I$ = (548.67 ± 0.015)+(239.83 ± 0.027)x, Fig. 5b). Notable, the interaction between ZnTPP and NO$_2$ is strong and irreversible, that the sensor shows no response to dry air after exposure to NO$_2$ gas.

The repetitive sensing behavior of as-fabricated sensors was evaluated by exposing alternatively to dry air and diluted NO$_2$ gases. At the experimental NO$_2$ concentrations (10 ppb - 100 ppm), the relative standard deviation (RSD) of the response values were small (Fig. 5c). As an example, for 100 ppm NO$_2$ gas, RSD value of intensity of output light was 3.45%. These results exhibited that the ZnTPP thin-film/K$^+$-exchanged glass OWG sensor are stable and repetitive that can be utilized for quantitative analysis of NO$_2$ gas. In addition, the sensing device had revealed greater response to NO$_2$ gas compared to H$_2$S, SO$_2$, HCl, and CO$_2$ gases at ambient temperature (Fig. S2).

The film thickness and aggregation degree of the sensitive ZnTPP molecules, which are related to the concentration of ZnTPP solution and rotating speeds of spin-coating, show distinct influence on the sensing performance (Fig. S2). The results confirm the ZnTPP thin-film/K$^+$-exchanged glass OWG sensing element with nanorod aggregations (prepared under the condition of mass concentration of 0.12% and rotating speed of 1050 rpm at 30 s) is the optimal one with highest sensitivity, i.e. show considerably largest response to NO$_2$ gas ($\Delta I$ value reached maximum change value up to 2194) compared with sensing ZnTPP thin-films prepared under other conditions.

### *UV-vis spectrum sensing*

Alternatively, to further investigate the sensing behavior of ZnTPP thin-film, we measured its UV-vis absorbance change under various concentrations (1 ppm to 100 ppm) of NO$_2$ gas. Fig. 5d show an observable absorbance change value ($\Delta Abs$) of 0.017 at wavelengths of 438 nm after exposure to 1 ppm of NO$_2$ gas, and the largest $\Delta Abs$ of 0.427 at wavelengths of 438 nm under 100 ppm of NO$_2$ gas. Moreover, the absorption peaks show no shift on neither Soret-band nor Q-bands. In order to compare the UV-vis sensitivity of the ZnTPP thin-film to ZnTPP thin-film/K$^+$-exchanged glass OWG sensing element under various concentrations of NO$_2$ gas, we listed all the response values transferred to SNRs in Table 1. UV-vis SNR of the ZnTPP thin-film to 1 ppm of

NO₂ are 29.4 and 0 at wavelengths of 438 nm and 650 nm, respectively. However, OWG SNR for ZnTPP sensor to 1 ppm of NO₂ is 109.6 at wavelengths of 650 nm, which is about 29.5 times higher than the UV-vis SNR at wavelengths of 438 nm. SNR for 5 ppm, 10 ppm, 50 ppm and 100 ppm of NO₂ gas to ZnTPP thin-film/K⁺-exchanged glass OWG sensors are about 14.3, 8.8, 5.3 and 2.2 times larger than the UV-vis Δ$Abs$ at wavelengths of 438 nm; and about unavailable, 787, 453.3, and 228.2 times higher than UV-vis at wavelengths of 650 nm, respectively. In principle, if the most sensitive wavelength at 438 nm laser beam utilized as a light source for the OWG gas sensor detection, it may be possible to monitor the NO₂ gas with lower concentration, even molecular level.

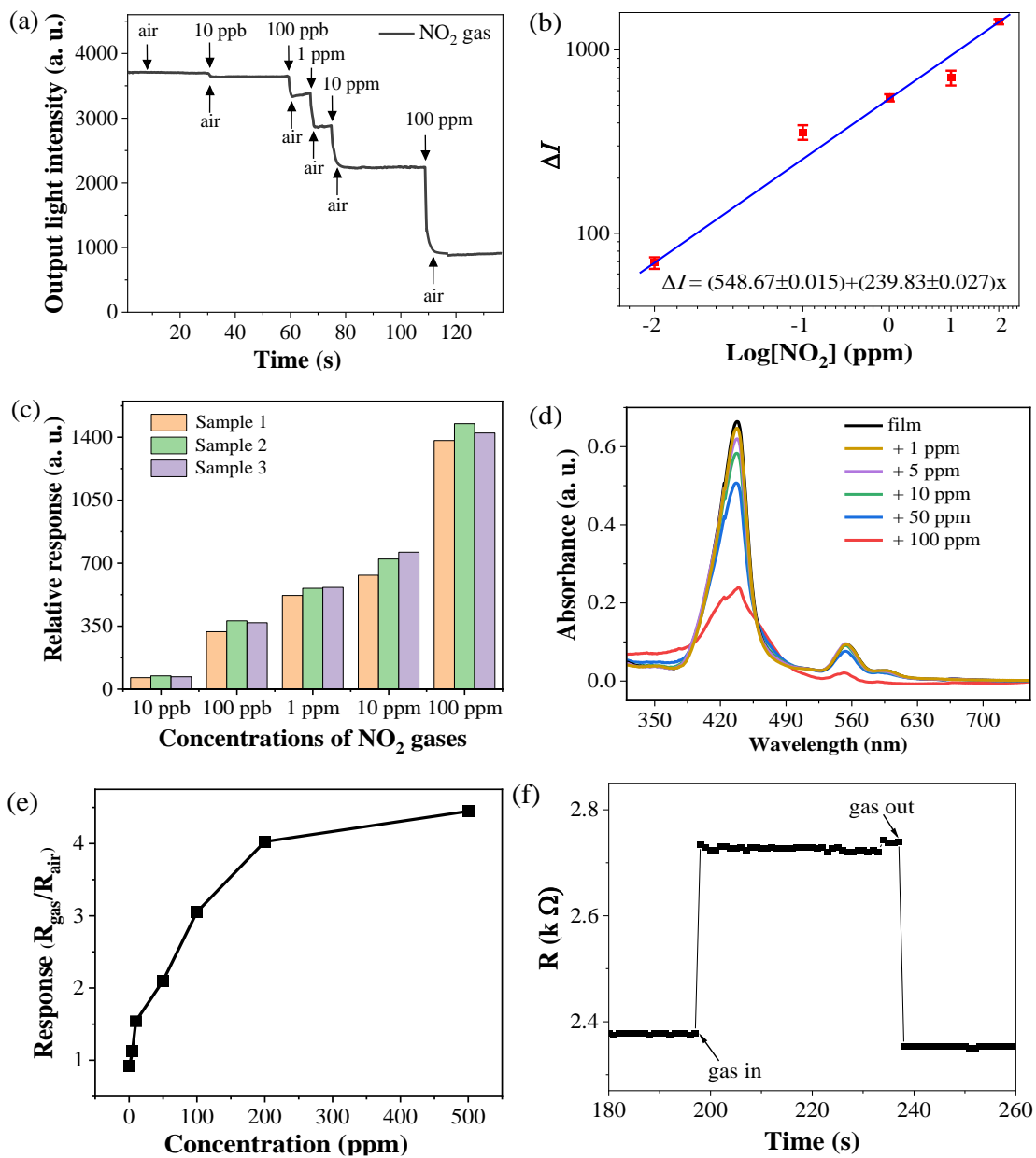

Fig. 5 Response curves of ZnTPP sensors to NO$_2$ gas. (a) Typical response of ZnTPP film/K$^+$-exchanged glass OWG sensor when exposed to NO$_2$ in air. (b) Linear relationship between the response values of the sensor and the concentrations of NO$_2$. (c) Repetitive response of ZnTPP film/K$^+$-exchanged glass OWG sensors to 10 ppb - 100 ppm of NO$_2$. (d) UV-vis sensitivity of ZnTPP films to NO$_2$ gas with different concentrations. (e) Response curve of ZnTPP electrical resistive sensor to different concentrations of NO$_2$ gases at room temperature. (f) Real time resistance changes of ZnTPP sensor to 1 ppm of NO$_2$ at room temperature.

Table 1. SNRs of OWG, UV-vis, and electrically operated gas sensors by utilizing ZnTPP as sensitizer upon exposure to NO$_2$ at different concentrations

| NO$_2$ gas concentration (ppm) | UV-vis | | OWG $\Delta I$ (650 nm) | Electrical response value ($R_{gas}/R_{air}$) |
|---|---|---|---|---|
| | $\Delta Abs$ (438 nm) | $\Delta Abs$ (650 nm) | | |
| 1 | 3.7 | 0 | 109.6 | 28 |
| 5 | 9.3 | 0 | 133.4 | 34.8 |
| 10 | 17.8 | 0.2 | 157.4 | 47.6 |
| 50 | 34.3 | 0.4 | 181.4 | 65.2 |
| 100 | 92.8 | 1.9 | 205.4 | 95.8 |

*Electrical resistance sensing*

For comparison, we further elucidated electronic resistance sensing performance of the ZnTPP film (dip-coated as Fig. 1b) to NO$_2$ gas. Fig. 5e shows the distinct electrical response of ZnTPP film after exposed to NO$_2$ gas of 1 ppm at room temperature. The resistive device demonstrates a high electrical response value ($R_{gas}/R_{air}$) of 0.93, 1.15, 1.57, 2.15, 3.16, 4.05 and 4.40 to NO$_2$ gas with concentration of 1 ppm, 5 ppm, 10 ppm, 50 ppm, 100 ppm, 200 ppm, and 500 ppm, respectively (Fig. 5e). Additionally, SNRs of the ZnTPP resistive devices are about 3.9, 3.8, 3.3, 2.8 and 2.2 times lower than that of the ZnTPP film/K$^+$-exchanged glass OWG sensor to NO$_2$ gas with concentration of 1 ppm, 5 ppm, 10 ppm, 50 ppm, and 100 ppm, respectively (Table 1). Notably, the resistance recovered after remove the NO$_2$ gas (Fig. 5f), which is differ from our observation with optical sensing methods. We suppose that due to oxygen molecules are absorbed on the surface of the ZnTPP electrical conductance sensor during aging process.

**Gas sensing mechanism**

In fact, metallo-porphyrins are electron-rich complex and $NO_2$ is a strong oxidizing gas. To probe into the chemical reaction mechanism between ZnTPP molecular and $NO_2$ gas, we mainly used FT-IR spectrum characterization together with UV-vis to discuss the detailed reaction process. Normally, ZnTPP can easily undergo coordination reaction with axial ligands (such as $NO_2$). As Fig. 6a shows, after exposure ZnTPP thin-film to $NO_2$ gas, asymmetric stretching band for nitrogen oxygen double bond ($v_{(N=O)}$) appears at 1644 $cm^{-1}$, while its strong symmetric stretching band appears at 1384 $cm^{-1}$. The FT-IR spectrum of the ZnTPP-$NO_2$ complex also exhibits peaks at 1582, 1442, and 844 $cm^{-1}$ that can be attributed to the out of plane bending vibration $\gamma_{\alpha(N=O)}$, $\gamma_{\beta(N=O)}$, and bending vibration $\delta_{(N=O)}$ of the ZnTPP-$NO_2$ complex, respectively.[27] These FT-IR spectrum peaks indicate the ZnTPP is oxidized by $NO_2$ and the formation of ZnTPP-$NO_2$ complex.

In addition, bending vibration peak $\gamma_{(N-Cm)}$ shifted from 996 $cm^{-1}$ to 998 $cm^{-1}$ and becomes widen, which proves the chemical bonding between ZnTPP and $NO_2$ molecules. Therefore, as the second reaction step another $NO_2$ nucleophilic radical is likely to attack meso-position of the porphyrin and reacts covalently with the oxidized porphyrin complex (Fig. 6b). Stimulates the UV-vis spectrum indicates distinct change and visible color of the ZnTPP thin-film changes from yellow to green (Fig. 4). Such combined spectral and visible color changes were also described in many papers.[28] Therefore, oxidation of ZnTPP by $NO_2$ in a two-step process (Fig. 6b) is the most probable mechanism for the sensing behavior.

It needs to be noted, for electrical resistance sensing the ZnTPP film needs to be aged under electrical power for 24 h. Similar to the metal oxide semiconductor[29, 30], the oxygen molecules in air are absorbed on the ZnTPP surface, and resulted electron-depletion layer and band bending on the surface along with the increasing resistance of the ZnTPP electrical conductance sensor. When an electron withdrawing gases (like $NO_2$) is introduced to ZnTPP film surface, chemical reactions take place between the gas molecules and the pre-absorbed oxygen ($O^-$ or $O^{2-}$), which gives out electrons withdraw from the semiconductor surface, and resulting in increased resistance of the sensing material.

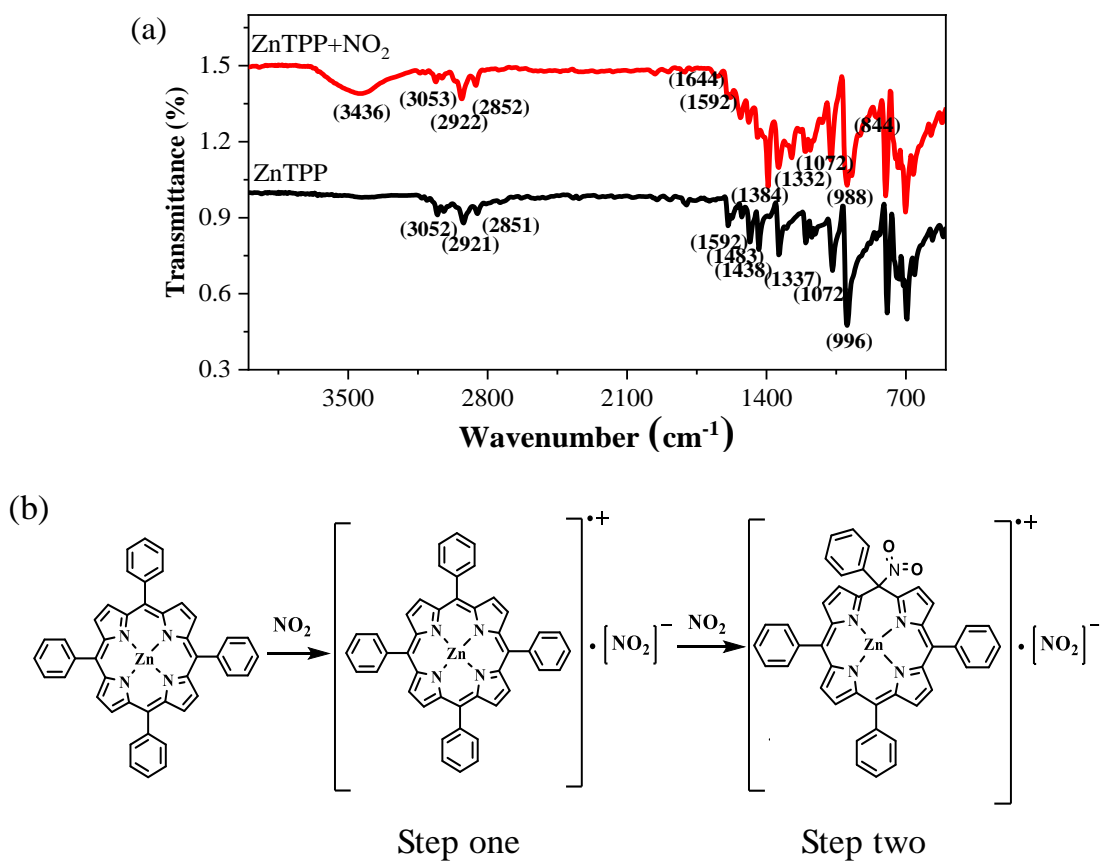

Fig. 6 Gas sensing mechanism of ZnTPP thin-film to $NO_2$ gas. (a) FT-IR spectra of ZnTPP thin-film before and after exposure to $NO_2$ gas, and (b) illustration of the two-step reaction process of ZnTPP complex with $NO_2$ gas.

**Conclusions**

A ZnTPP based $NO_2$ gas sensor has been developed and evaluated by OWG technique, UV-vis spectrum and electrical resistance measurement. ZnTPP thin-film/$K^+$-exchanged glass OWG sensing device could detect very low concentration of concentration of $NO_2$ to 10 ppb with SNR of 3.4, which is undetectable by UV-vis method. SNRs comparation under the same $NO_2$ concentrations indicate the ZnTPP thin-film/$K^+$-exchanged glass OWG element has higher sensitivity than both the UV-vis spectrum and electrical resistance measurement. The reaction mechanism for optical based $NO_2$ sensing method and electrical resistance sensing for ZnTPP were discussed by oxidation-reduction reaction, respectively. Compared with metal oxide semiconducting sensors, the optimal prepared ZnTPP thin-film/$K^+$-exchanged glass OWG sensor has advantages of easy to preparation, simple structure, highly selective, and operating at room temperature, which is possible to pave a new way for the exploration of optical sensing devices.

Supplementary Information

# Exploring Optical and Electrical Gas Detection Based on Zinc-Tetraphenylporphyrin Sensitizer

Gulimire Tuerdi*, Abliz Yimit**, Xiaoyan Zhang*†

*School of Materials Science and Engineering, Tsinghua University, Beijing 100084, China.
**College of Chemistry and Chemical Engineering, Xinjiang University, Urumqi 830046, China.
†To whom correspondence should be addressed
E-mail: xyzhang@iccas.ac.cn

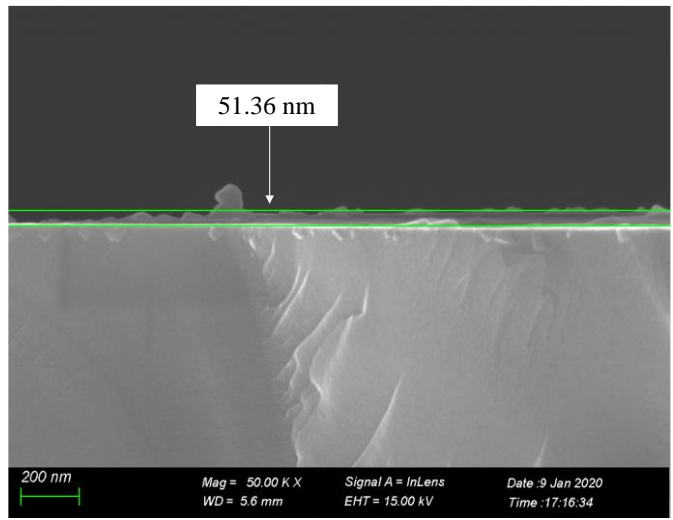

Fig. S1 Film thickness of ZnTPP film OWG gas sensing element measured by SEM.

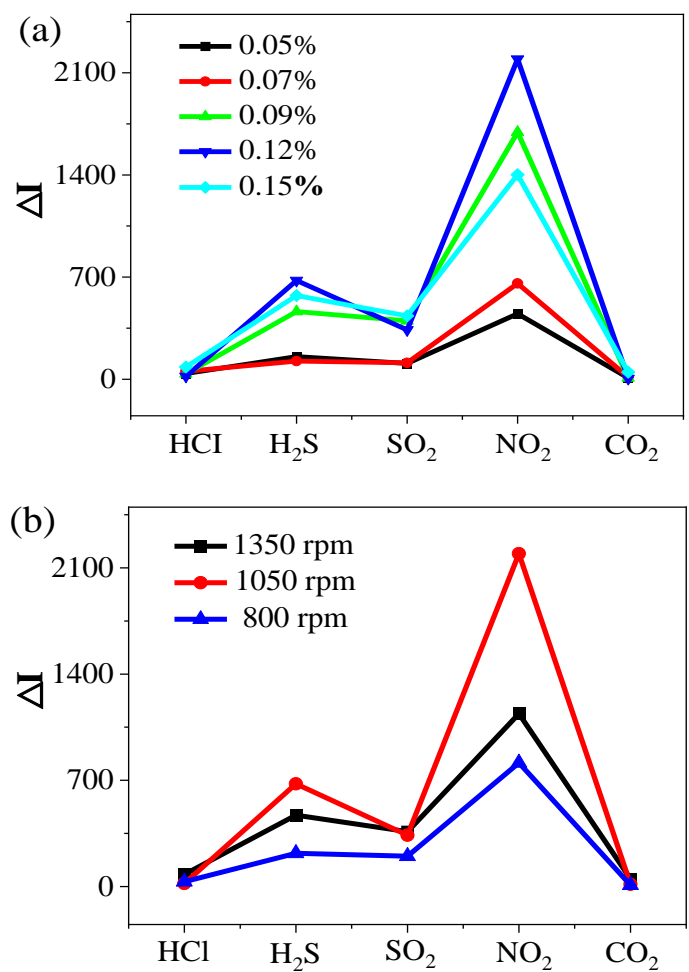

Fig. S2 Typical response of ZnTPP thin-film OWG sensor to different inorganic gases (a) under different mass concentration solutions and (b) different rotating speeds.